\documentclass[aps,twocolumn]{revtex4}
\usepackage{epsfig}
\usepackage{bm}
\def\B.#1{{\bm{#1}}}
\def\C.#1{{\cal{#1}}}

\def\i{{\rm i}}
\def\d{{\rm d}}

\def\ep{{\epsilon}}
\def\be{\begin{equation}}
\def\ee{\end{equation}}

\def\R{{\rm Re}} 
\def\k{k} 

\def\s{\left(k_+^2 + \beta^2\right)}
\def\A{\left[\i (\omega_+ - k_+ U) + \nu(D^2 - k_+^2 - \beta^2) +(D\nu)
D\right]}

\def\beta{k_z}

\begin{document} 

\title{Stabilization of Hydrodynamic Flows by Small Viscosity Variations}
\author{Rama Govindarajan$^{\dag}$, Victor S. L'vov$^*$ and Itamar
Procaccia$^*$}
\affiliation{$^\dag $ Fluid Dynamics Unit, Jawaharlal Nehru Centre for
Advanced 
Scientific Research, Jakkur, Bangalore 560064, India. \\
$^*$ Dept. of Chemical Physics, The Weizmann Institute of Science, Rehovot
76100, Israel. }
 
\begin{abstract}  
Motivated by the large effect of turbulent drag reduction by minute
concentrations of polymers we study the effects of a weakly
space-dependent viscosity on the stability of hydrodynamic flows. In a
recent Letter [Phys. Rev. Lett. {\bf 87}, 174501, (2001)] we exposed
the crucial role played by a localized region where the energy of
fluctuations is produced by interactions with the mean flow (the
``critical layer"). We showed that a layer of weakly space-dependent
viscosity placed near the critical layer can have a very large
stabilizing effect on hydrodynamic fluctuations, retarding
significantly the onset of turbulence.  In this paper we extend these
observation in two directions: first we show that the strong
stabilization of the primary instability is also obtained when the
viscosity profile is realistic (inferred from simulations of turbulent
flows with a small concentration of polymers).  Second, we analyze the
secondary instability (around the time-dependent primary instability)
and find similar strong stabilization. Since the secondary instability
develops around a time-dependent solution and is three-dimensional,
this brings us closer to the turbulent case. We reiterate that the
large effect is {\em not} due to a modified dissipation (as is assumed
in some theories of drag reduction), but due to reduced energy intake
from the mean flow to the fluctuations. We propose that similar
physics act in turbulent drag reduction.
\end{abstract}

\maketitle

\section{Introduction}
 
This paper is motivated by the dramatic effects that are observed with
the addition of small amounts of polymers to hydrodynamic flows.
While interesting effects were discussed in the context of the
transition to turbulence, vortex formation and turbulent transport
\cite{95NH}, the phenomenon that attracted the most attention was, for
obvious reasons, the reduction of friction drag by up to 80\% when
very small concentrations of long-chain polymers were added to
turbulent flows \cite{69Lum,00SW}. In spite of the fact that the
phenomenon is robust and the effect huge, there exists no accepted
theory that can claim quantitative agreement with the experimental
facts. Moreover, it appears that there is no mechanistic
explanation. In the current theory that is due to de Gennes
\cite{86TG,90Gennes} one expects the Kolmogorov cascade to be
terminated at scales larger than Kolmogorov scale, leading somehow to
an increased buffer layer thickness and reduced drag, but how this
happens and what is the fate of the turbulent energy is not being made
clear.

\begin{figure} 
\epsfxsize=7.5cm \epsfbox{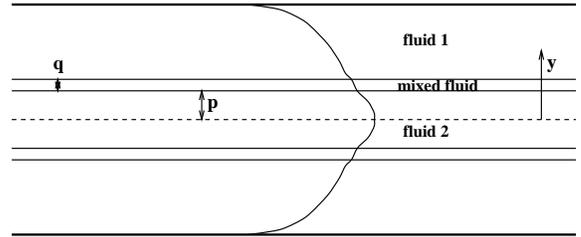}
\caption{Schematic of the flow: the fluid near the walls has a viscosity
$\mu_1$, and that flowing at the center is of viscosity $\mu_2$. In
the mixed layer (of width $q$) the viscosity varies gradually between
$\mu_1$ and $\mu_2$. The parameter $p$ controls the position of the
mixed layer. For simplicity we neglect the down-stream growth in $q$.}
\label{f:scheme} 
\end{figure}

In a recent Letter \cite{01GLP} we proposed that the crucial issue is
in the {\em production} of energy of hydrodynamic fluctuations by
their interaction with the mean flow. For the sake of concreteness we
examined a Poiseuille laminar flow and its loss of linear stability,
and showed how small viscosity contrasts lead to an order of magnitude
retardation in the onset of instability of ``dangerous"
disturbances. Specifically, we considered a flow in a channel of
dimensionless width 2, in which there are two fluids: one fluid of
viscosity $\mu_1$ flows near the walls, and the other fluid of
viscosity $\mu_2$ flows at the center, see Fig. \ref{f:scheme}. The
viscosities differ slightly, for example we considered (in
dimensionless units) $\mu_2=1$ and $m=\mu_1/\mu_2 =0.9$. The main
ingredient of the calculation was that all the viscosity difference of
0.1 concentrated in a ``mixed" layer of width 0.10. The motivation
behind these numbers was the observation that the inferred effective
viscosity in polymer drag reduction increases towards the center by
about 30\% over about a 1/3 of the half-channel \cite{97SBH}. With our
choice we have comparable viscosity gradients in the mixed layer.

In this model everything was explicitly calculable. The main point of
our analysis (see Sect. \ref{primary} for further details) was that
there exists a position in the channel where the velocity of the mean
flow is the same as the velocity of the most dangerous primary
instability. Below we refer to the layer around this position as the
``critical layer". If we placed the mixed layer in the vicinity of the
critical layer, we got a giant effect of stabilization.  Analyzing
this phenomenon, we demonstrated that nothing special happened to the
dissipation.  Rather, it was the energy intake from the mean flow to
the unstable mode that was dramatically reduced, giving rise to a
large effect for a small cause. In this paper we extend these
observations in two directions. In Sect. \ref{primary}, after
reviewing the results of the simple model, we extend the analysis of
the primary instability to a case in which the viscosity profile is
that inferred from direct numerical simulations of turbulent channel
flow of a dilute polymeric solution \cite{97SBH}. We will see that
very similar effects are found. In other words, one does not need to
put by hand the region of viscosity variation in the vicinity of the
critical layer. When we have a continuous variation of the viscosity
in the region near the wall, the effect is the same, since it is only
crucial that there will be {\em some} space dependence of the
viscosity in the critical layer, which is usually not too far from the
wall.

A possible criticism of our results can be that the primary
instability is still too far from typical turbulent fluctuations. This
is in particular true since the most unstable primary modes are
2-dimensional, whereas typical turbulent fluctuations are
3-dimensional. For these reasons we present in Sect. \ref{secondary}
the analysis of the effect of small viscosity variations on the
secondary instability, for which the most ``dangerous" modes are
3-dimensional. The tactics are similar to those taken for the primary
instability. First we discuss the effects of a mixed layer put at the
``right" place in the channel, and second we show that continuous
viscosity profiles do exactly the same. We find again the giant effect
of stabilization for relatively small viscosity variations, lending
further support to our proposition that similar effects may very well
play a crucial role in turbulent drag reduction.  In Sect.
\ref{conclude} we present concluding remarks and suggestions for the
road ahead.

\section{Primary instability of Poiseuille flow}
\label{primary}

It is well known that parallel Poiseuille flow loses linear stability
at some threshold Reynolds number Re=Re$_{\rm th}$ (close to 5772).
It is also well known that the instability is ``convective", with the
most unstable mode having a phase velocity $c_p$. Analytically it has
the form \be \hat \phi_p(x,y,t) = \frac{1}{2}\big\{\phi_p(y) \exp\left
[\i \k_p(x-c_p \, t) \right ] + \mbox{c.c.}\big\} \exp(\gamma_p t) \ ,
\label{eq:inst} 
\ee where the subscript $p$ stands for the primary instability, $\hat
\phi(x,y,t)$ is the disturbance streamfunction and $\phi(y)$ is the
complex envelope of $\hat \phi(x,y,t)$. We have chosen $x$ and $y$ as
the streamwise and wall-normal coordinates respectively, $\k$ as the
streamwise wavenumber of the disturbance and $t$ as time. $\gamma_p$
is the growth rate of the primary instability.  What is not usually
emphasized is that the main interactions leading to the loss of
stability occur in a sharply defined region in the channel, i.e. at
the critical layer whose distance from the wall is such that the phase
velocity $c$ is identical to the velocity of the mean flow somewhere
within this layer. It is thus worthwhile to examine the effect on the
stability of Poiseuille flow of a viscosity gradient placed in the
vicinity of the critical layer. This will provide us with a very sharp
understanding of the mechanism of the stabilization of the flow by
viscosity variations. In the following subsection we will examine the
case of continuous viscosity profiles.

\subsection{Mixed Layer}
\label{primix}

A report of the results of this subsection was provided in
\cite{01GLP}. We examine a channel flow of two fluids with different
viscosities $\mu_1$ and $\mu_2$, see Fig. \ref{f:scheme}.

\begin{figure}
\epsfxsize=8.3
cm
\epsfbox{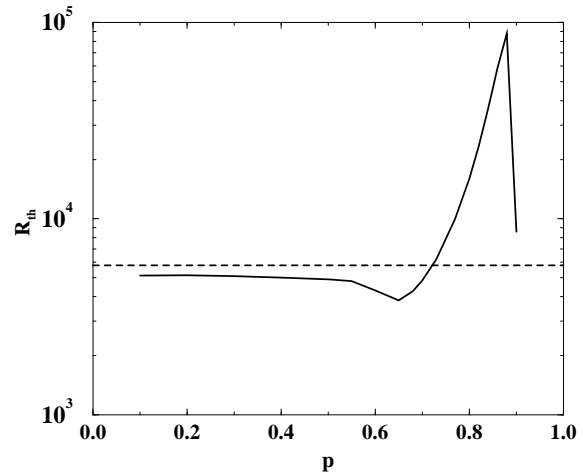}
\caption{The dependence of the threshold Reynolds
number Re$_{\rm th}$ on the position of the viscosity stratified layer
for $m=0.9$. The dashed line pertains to the neat fluid. Note the huge
increase in $\Re_{\rm th}$ within a small range. This occurs when the
stratified layer overlaps the critical layer.}
\label{f:rcr}
\end{figure}

The observation that we want to focus on is shown in Fig. \ref{f:rcr}:
the threshold Reynolds number for the loss of stability of the mode as
in Eq.  (\ref{eq:inst}) depends crucially on the position of the mixed
layer. When the latter hits the critical layer, the threshold Reynolds
number for the loss of stability reaches as much as 88000. In other
words, one can increase the threshold of instability {\em for a given
mode} 15 times, and by making the mixed layer thinner one can reach
even higher threshold Reynolds values.
In \cite{01GLP} we analyzed the physical origin of this huge
sensitivity of the flow stability to the profile of the viscosity.

The stability of this flow is governed by the modified Orr-Sommerfeld
equation \cite{white}
\begin{eqnarray}
\nonumber
&& \i\k_p
\left[\left(\phi_p''-\k_p^2\phi_p\right)(\bar U-c_p-\i
\gamma_p) - \bar
U''\phi_p \right] \nonumber\\&&= {1\over {\rm Re}}\bigg[\mu
\phi_p^{\rm (4)}
+ 2
\mu' \phi_p''' +\left(\mu'' - 2 \k_p^2 \mu \right)\phi_p''
\nonumber\\&&-
2
\k_p^2 \mu' \phi_p' +
\left(\k_p^2 \mu'' + \k_p^4 \mu\right)
\phi_p\bigg]\,,
\label{modOS} 
\end{eqnarray} 
in which $\bar U(y)$ is the basic laminar velocity, and $\mu$ is a
function of $y$. The boundary conditions are $\phi_p(\pm 1) = \phi_p'
(\pm 1) = 0$. All quantities have been non-dimensionalised using the
half-width $H$ of the channel and the centerline velocity $U_0$ as the
length and velocity scales respectively.  The Reynolds number is
defined as Re$\equiv \rho U_0 H /\mu_2$, where $\rho$ is the density
(equal for the two fluids). The primes stand for derivative with
respect to $y$.  At $y=0$, we use the even symmetry conditions
$\phi(0)=1$ and $\phi'(0)=0$, as the even mode is always more unstable
than the odd.

\begin{figure}
\epsfxsize=7.5cm  \epsfbox{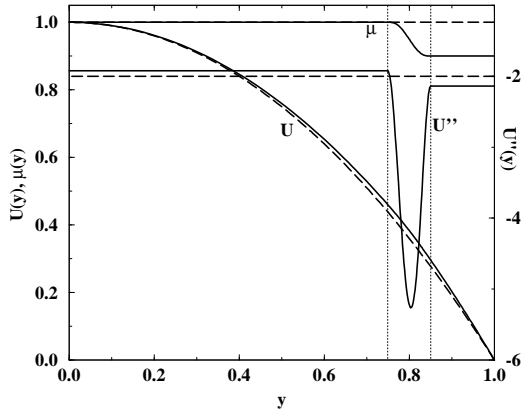}
\caption{Profiles of the normalized
viscosity $\mu(y)$ and normalized velocity $\bar U(y)$ and the second
derivative $\bar U^{''}(y)$ for $m=0.9$ (solid lines) and $m=1.0$
(dashed lines). The mixed layer is between the vertical dashed lines.
}
\label{f:thinprof} 
\end{figure}

Since the flow is symmetric with respect to the channel centreline, we
restrict our attention to the upper half-channel. Fluid 2 occupies the
region $0 \le y \le p$. Fluid 1 lies between $p+q \le y \le 1$. The
region $p \le y \le p+q$ contains mixed fluid. The viscosity is
described by a steady function of $y$, scaled by the inner fluid
viscosity $\mu_2$:
\begin{eqnarray}\label{muin}
\hskip -0.4cm \mu(y) &=& 1
\,,\quad  \mbox{for} \quad 0 \le y \le p\,,\\
\label{vis5}
\hskip -0.4cm \mu(y) &=& 1 +
(m-1)\, \xi^3\left[10 - 15\, \xi + 6 \xi^2
\right], \ 0 \le \xi \le 1\,,
\\
\label{muout}
\hskip -0.4cm \mu(y) &=& m \,,\quad  \mbox{for} \quad p+q \le y \le
1.
\end{eqnarray}
Here $\xi\equiv (y-p)/q$ is the mixed layer coordinate.  We have
assumed a 5th-order polynomial profile for the viscosity in the mixed
layer, whose coefficients maintain the viscosity and its first two
derivatives continuous across the mixed layer. The exact form of the
profile is unimportant. For a plot of the profile $m=0.9$, see
Fig.~\ref{f:thinprof}.

The basic flow $\bar U(y)$ is obtained by requiring the velocity and
all relevant derivatives to be continuous at the edges of the mixed
layer:
\begin{eqnarray}\label{ui}
\hskip -0.5cm &&   \bar U(y) = 1-Gy^2/2\,, \quad
\mbox{for} \quad y \le p\,,\\
\hskip -0.5cm &&\bar U(y) = U(p) - G\int_p^ydy~ y/ \mu
\,,\quad  \mbox{for}
\quad p \le y \le p+ q,
\label{um}\\
\hskip -0.5cm && \bar U(y) =
G\left(1-y^2\right)/2m, \quad  \mbox{for}
\quad y \ge p+q\
.\label{uo}
\end{eqnarray}
Here $G$ is the streamwise pressure gradient.

It can be seen, comparing the mean profile $\bar U(y)$ to that of the
neat fluid (cf. Fig. \ref{f:thinprof}), that nothing dramatic happens
to this profile even when the mixed layer is chosen to overlap a
typical critical layer. Accordingly we need to look for the origin of
the large effect of Fig. 2 in the energetics of the disturbances. To
do so, recall that the streamwise and normal components of the
disturbance velocity $\hat u_p(x,y,t)$ and $\hat v_p(x,y,t)$ may be
expressed via streamfunction as usual: $\hat u_p(x,y,t) = \partial\hat
\phi_p/\partial y$, ${\rm and} \quad \hat v_p(x,y,t) = -\partial \hat
\phi_p/\partial x$.  These functions may be written in terms of
complex envelopes similar to Eq.~(\ref{eq:inst}):
\begin{eqnarray} 
  \label{eq:vel-env}
  \hat u_p(x,y,t) &\!\!\!=\!\!\!& \frac{1}{2}\big\{u_p(y)
  \exp\left[\i\k_p(x\!-\!c_p\, t)\right]\!+\!  \mbox{c.c.}\big\}
  \exp(\gamma_p t),\\ \nonumber \hat v_p(x,y,t) &\!\!=\!\!&
  \frac{1}{2}\big\{v_p(y) \exp\left[\i\k_p(x-c_p\, t)\right] +
  \mbox{c.c.}\big\} \exp(\gamma_p t) \ .
\end{eqnarray} 
The pressure
disturbance $\hat p_p$ is defined similarly.

Define now a disturbance of the density of the kinetic ene rgy of the
primary instability 
\be\label{kinen} \hat E_p(x,y,t) =
\frac{1}{2}\left[ \hat u_p(x,y,t)^2 + \hat v_p(x,y,t)^2\right] \ .
\ee We can express the mean (over $x$) density of the kinetic energy
as follows:
\begin{eqnarray}\label{averen}
  E_p(y,t)&\equiv& \left< \hat
E_p(x,y,t)\right>_x =\C.E_p (y)\exp\,(2
\gamma_p t)\,,
\\ \nonumber
\C.E_p
(y)&=&\frac{1}{4}\left( |u_p(y)|^2 + |v_p(y)|^2 \right)\ .
\end{eqnarray}

The physics of our phenomenon will be discussed in terms of the
balance equation for the averaged disturbance kinetic energy. Starting
from the linearized Navier-Stokes equations for $\hat u_p$ and $\hat
v_p$, dotting it with the disturbance velocity vector, averaging over
one cycle in $x$ and using Eqs. (\ref{eq:vel-env})-(\ref{averen})
leads to \be 2 \gamma_p \,\C.E_p (y) = \nabla \cdot J_p(y) + W_{p+}(y)
- W_{p-}(y) \,,
\label{enbal} 
\ee where the energy flux $J_p(y)$ in the $y$ direction, rates of
energy production (energy taken up by the primary instability from the
mean flow) $W_{p+}(y)$ and energy dissipation (by the viscosity)
$W_{p-}(y)$ are given by
\begin{eqnarray}\label{current}
  J_p(y) &\equiv&
{\left[u_p(y) p_p^*(y)+\mbox{c.c.}\right]
\over 4 \rho} + {1 \over \R} \mu
(y) \nabla \C.E_p(y)\,,\\
W_{p+}(y) &\equiv&  - {1 \over 4} \bar
U'(y)\left[u_p(y)
 v_p^*(y) +\mbox{c.c.}  \right]\,,
\label{prod}
\\
W_{p-}(y) &\equiv& {\mu(y) \over \R} \left\{2 \k_p^2 \C.E_p(y)
+ {1 \over
2}\left[|u_p'(y)|^2+|v_p'(y)|^2\right]\right\}\
.
\nonumber
\end{eqnarray}
The superscript $*$ denotes complex conjugate. To plot these functions
we need to solve Eq. (\ref{modOS}) as an eigenvalue problem, to obtain
$c_p$, $\gamma_p$, and $\phi_p(y)$ at given $\R$ and $\k_p$.  The
value of $c_p$ determines the position of the critical layer. It is
convenient to compute and compare the space averaged production and
dissipation terms $\Gamma_{p+}$ and $\Gamma_{p-}$ defined by:
\be\label{totpd} \Gamma_{p\pm} \equiv\int_0^1 W_{p\pm}(y) \d y \ \Big/
\int_0^1 \C.E_p(y) \d y \ .  \ee The local production of energy can be
positive or negative, indicative of energy transfer from the mean flow
to the primary disturbance and vice-versa respectively. The production
in one region (where $ W_{p+}(y)>0$) can be partly canceled out by a
``counter-production'' in other region (where $ W_{p+}(y)<0$).

\begin{figure} 
\epsfxsize=7.5cm
\epsfbox{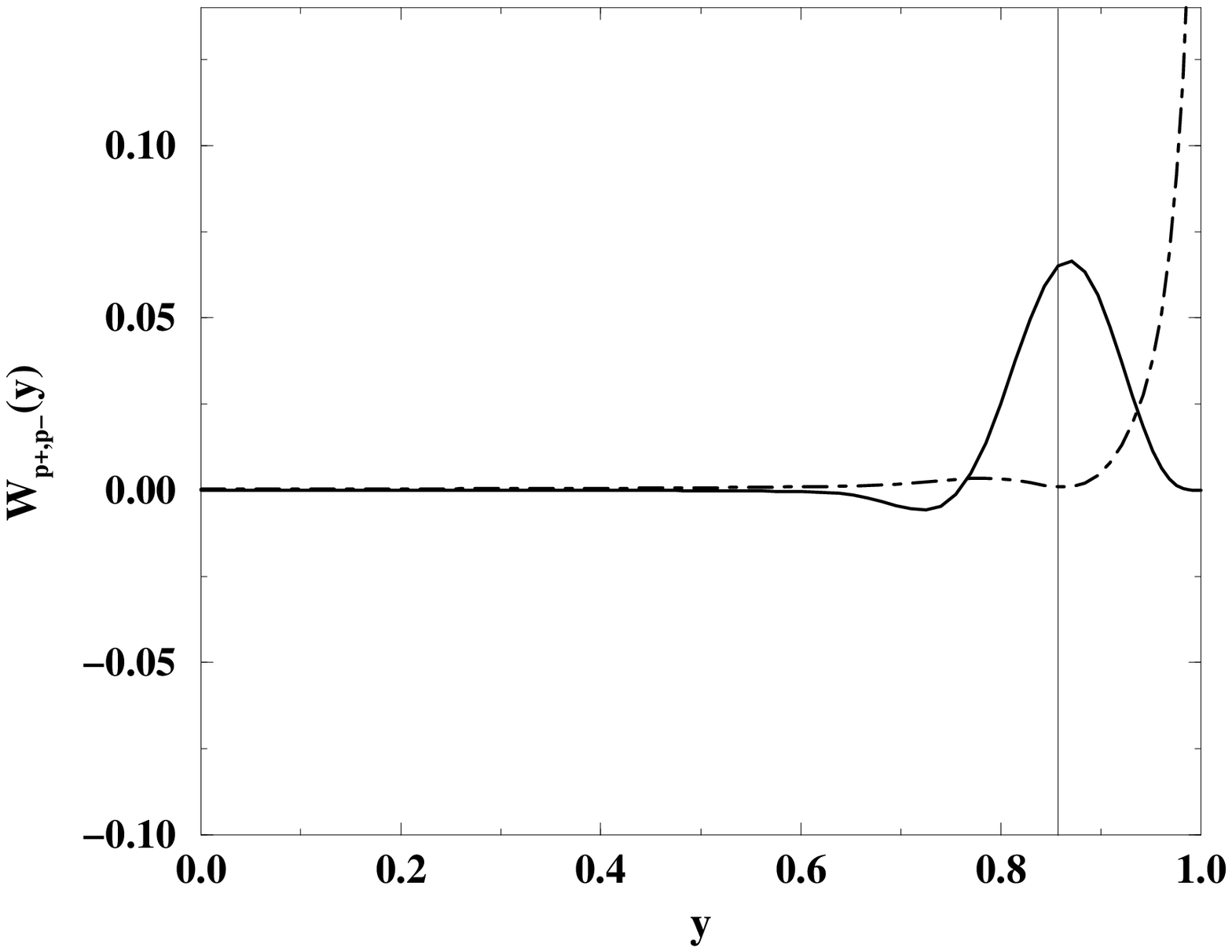}
\epsfxsize=7.5cm \epsfbox{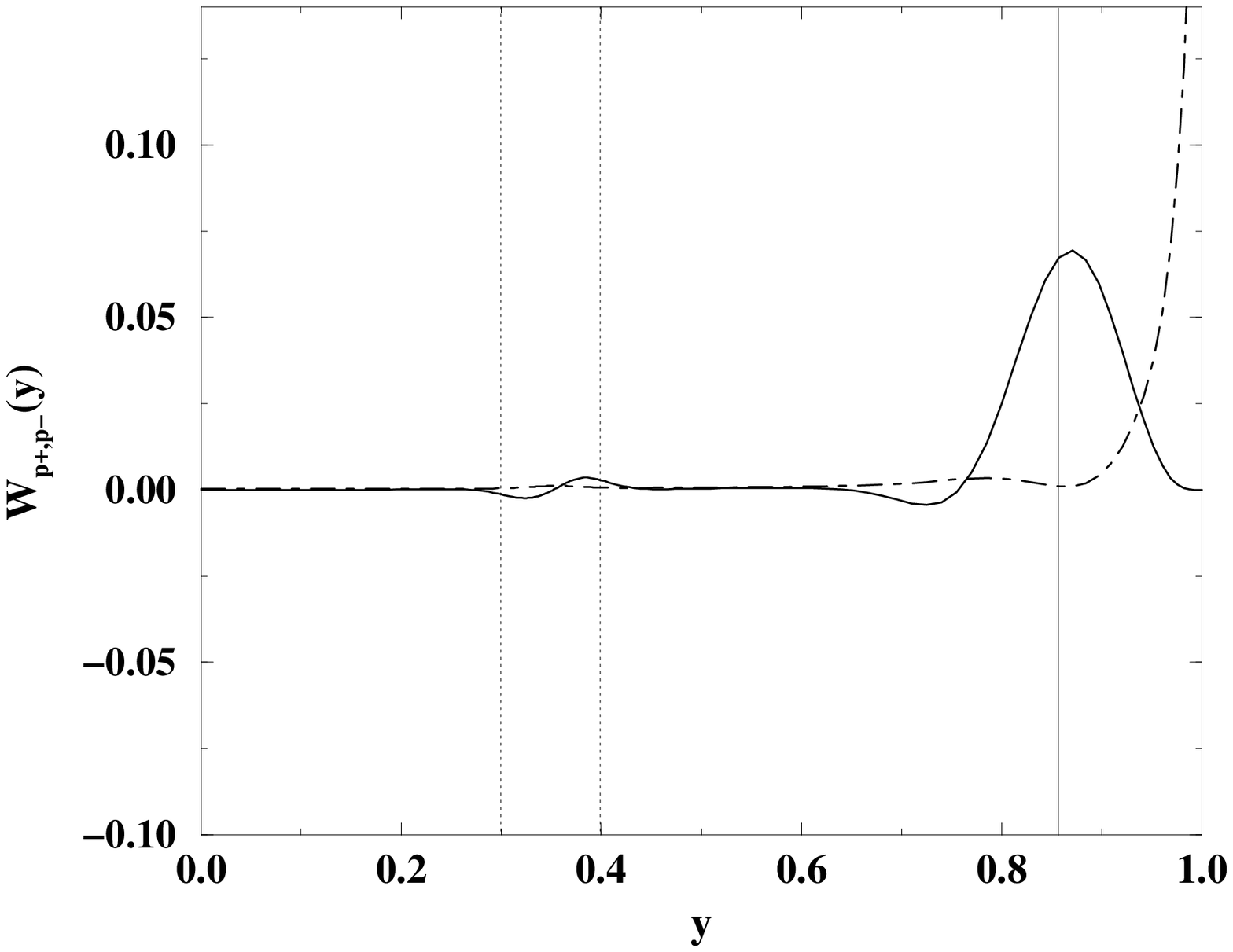}
\caption{Energy
balance: production $W_{p+}(y)$, solid line; dissipation
$W_{p-}(y)$,
dot-dashed line, $\R=5772$. Top: $m=1$, $\Gamma_{p+}=
\Gamma_{p-}=0.0148$.
Bottom: $m=0.9$, $p=0.3$, $\Gamma_{p+}=0.0158$,
$\Gamma_{p-}=0.0148$. In
this and the two subsequent figures the solid
vertical lines show the
location $y_c$ of the critical lines, whereas the
region
between the dotted
lines is the mixed layer.}
\label{f:far} 
\end{figure}

The use of these measures can be exemplified with the neat fluid
($m=1.0$ here). The laminar flow displays its first linear instability
at a threshold Reynolds number of $\R_{\rm th}=5772$, which means that
the total production $\Gamma_{p+}$ across the layer becomes equal to
the total dissipation $\Gamma_{p-}$ at this value of $\R$. Examining
Fig.  \ref{f:far} we can see that the disturbance kinetic energy is
produced predominantly within the critical layer, where the basic flow
velocity is close to the phase speed of the disturbance, while most of
the dissipation is in the wall layer.
The balance is not changed significantly when the viscosity ratio is
changed to $0.9$ so long as the mixed layer is not close to the
critical layer.  There is a small region of production and one of
counter-production within the mixed layer, whose effects cancel out,
leaving the system close to marginal stability.
\begin{figure}
\epsfxsize=7.5cm \epsfbox{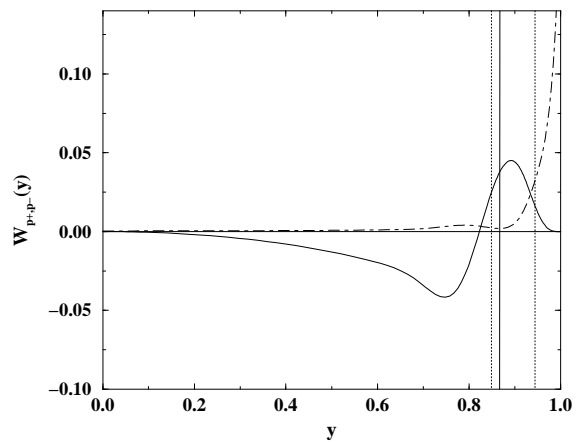}
\caption{Energy balance: production
$W_{p+}(y)$, solid line; dissipation
$W_{p-}(y)$, dot-dashed line.
$\R=5772$, $m=0.9$, $p=0.85$, $\Gamma_{p+}=
-0.0114$, $\Gamma_{p-}=0.0122$.
}
 \label{f:near} 
\end{figure} 

\begin{figure}
\epsfxsize=7.5cm
\epsfbox{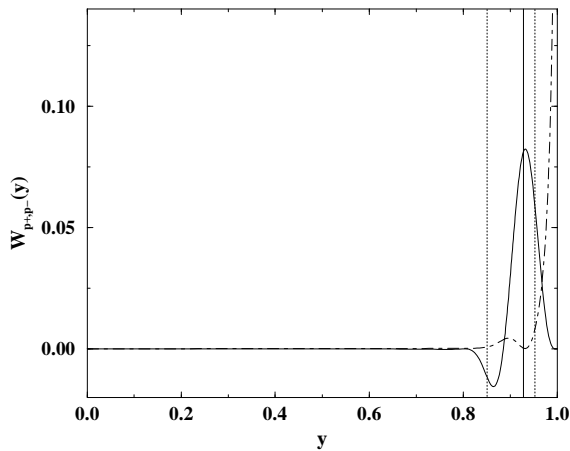}
\caption{Energy balance: production $W_{p+}(y)$,
solid line; dissipation
$W_{p-}(y)$, dot-dashed line. $\R=46400$, $m=0.9$,
$p=0.85$, $\Gamma_{p+}=
\Gamma_{p-}=0.0053$. }
 \label{f:marg} 
\end{figure}

We now turn our attention to Fig. \ref{f:near}, in which our main
point is demonstrated.  The Reynolds number is the same as before, but
the mixed layer has been moved close to the critical layer. It is
immediately obvious that the earlier balance is destroyed. The
counter-production peak in the mixed layer is much larger than before,
making the flow more stable.  The wavenumber used is that at which the
flow is least stable for the given Reynolds number at this $p$. For
$m=0.9$, the threshold Reynolds number is $46400$.  Fig.  \ref{f:marg}
shows the energy balances at marginal stability - the picture is
qualitatively the same here as at $\R\approx 5772$ for the neat fluid.
\subsection{The mechanism of
stabilization}
 
The main factor determining the instability is the energy intake from
the mean flow, which is driven by the phase change caused by the
viscosity stratification. The dissipation on the other hand depends
only on Reynolds number and does not respond disproportionately to
changes in viscosity. In neat fluids, the term containing $\bar
U''(y)$ in (\ref{modOS}) is always of higher order within the critical
layer. However, with the introduction of a viscosity gradient within
the critical layer, the gradients of the basic velocity profile will
scale according to the mixed layer coordinate $\xi$. We show in the
analysis that follows that for $q \le O(\R^{-1/3})$, the term
containing $\bar U''$ is now among the most dominant. Since most of
the production of disturbance kinetic energy takes place within the
critical layer, we return to equation (\ref{modOS}) and isolate all
lowest-order effects within the critical layer. The relevant normal
coordinate in the critical layer is \be \eta \equiv {y-y_c \over \ep}
\label{eta}
\ee
where $y_c$ is the critical point defined by $U(y_c)=c$, and $\ep$ is the
critical layer thickness, which is a small parameter at large Reynolds
number. The basic channel flow velocity may be expanded in the vicinity of
the critical point as
\be
U(y) = c + (y-y_c) U'(y_c) + {(y-y_c)^2 \over 2!} U''(y_c) + \cdots.
\label{ucrit}
\ee
We use (\ref{ucrit}), and redefine $\phi_p(y)\equiv\Phi(\eta)$ and
$\mu(y)\equiv \nu(\xi)$, to rewrite
(\ref{modOS}) within the critical layer. We obtain
\be
\ep \sim \R^{-1/3} \equiv \left(\k_p\R\right)^{-1/3},
\label{epsilon}
\ee and the lowest order equation in the critical layer: \be \i \eta
{\d U \over \d y}\bigg|_c\Phi'' - {\i G p \over \nu^2} \chi\nu'\Phi =
\nu \Phi^{\rm (4)} + 2 \chi\nu' \Phi''' + \chi^2\nu''\Phi''\,,
\label{critlow}
\ee where $\chi \equiv \ep /q$ is $O(1)$ for the mixed layer. In the
absence of a viscosity gradient in the critical layer (i.e. $\nu=1$),
equation (\ref{critlow}) would reduce to \be \i \eta {\d U \over \d
y}\bigg|_c\Phi'' = \Phi^{\rm (4)}\,,
\label{lin}
\ee which is the traditional lowest-order critical layer equation for
a parallel shear flow \cite{45Lin}. The mechanism for the
stabilization now begins to be apparent: there are several new terms
which can upset the traditional balance between inertial and viscous
forces. In order to narrow down the search further, we resort to
numerical experimentation, because although all terms in
(\ref{critlow}) are estimated to be of $O(1)$, their numerical
contributions are different.  It transpires that the second term on
the left hand side of (\ref{critlow}) is particularly responsible: it
is straightforward to verify that it originates from the term
containing $\bar U''(y)$ in the modified Orr-Sommerfeld equation. As
testimony, note the dramatic effect on $\bar U''$ in Fig. 3. Any
reasonable viscosity gradient of the right sign will pick up this
term, leading to vastly enhanced stability.

Indeed, in the light of this discussion we can expect that the large
effect of retardation of the instability would even increase if we
make the mixed layer thinner. This is indeed so. Nevertheless, one
cannot conclude that instability can be retarded at will, since other
disturbances, differing from the primary mode, become unstable first,
albeit at a much higher Reynolds number than the primary mode; when we
stabilize a given mode substantially, we should watch out for other
pre-existing/newly destabilized modes which may now be the least
stable.

Finally, we connect our findings to the phenomenon of drag reduction
in turbulent flows. Since the total dissipation can be computed just
from the knowledge of the velocity profile at the walls, any amount of
drag reduction must be reflected by a corresponding reduction of the
gradient at the walls.  Concurrently, the energy intake by the
fluctuations from the mean flow should reduce as well. Indeed, the
latter effect was measured in both experiments \cite{97THKN} and
simulations \cite{01DSBH,00Ang}. The question is which is the chicken
and which is the egg. In our calculation we identified that the
reduction in production comes first. From Figs. 4 and 5 which are at
the same value of $\R$ we see that the dissipation does not change at
all when the mixed layer moves, but the production is strongly
affected. Of course, at steady state the velocity gradient at the wall
must adjust as shown in Fig. 6.

\subsection{Continuous Viscosity
Profile}
\label{contprim}

\begin{figure}
\epsfxsize=7.5cm
\epsfbox{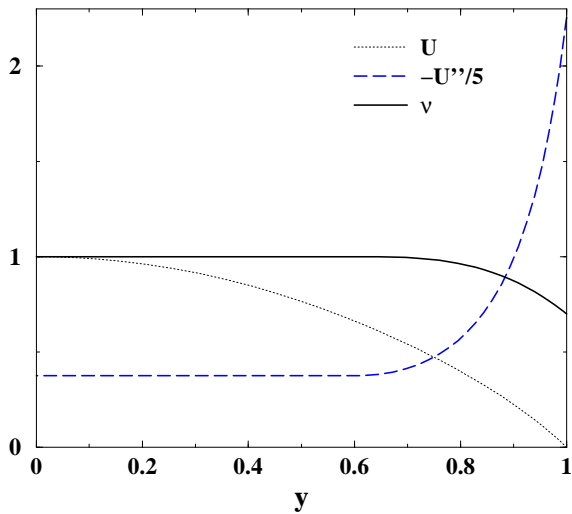}
\caption{Prescribed continuous viscosity profile
(in accordance with that
obtained in direct numerical simulations of
polymeric flow). The
corresponding laminar velocity profile $\bar U(y)$ and
its second derivative
are also shown.}
\label{f:contprof}
\end{figure}

\begin{figure} 
\epsfxsize=7.5cm \epsfbox{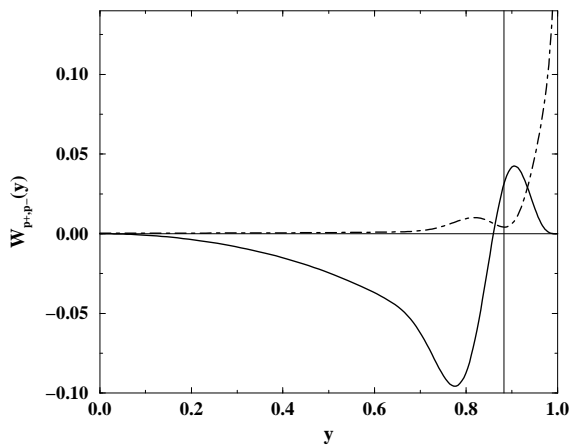}
\caption{Energy balance: production $W_{p+}(y)$, solid line; dissipation
$W_{p-}(y)$, dot-dashed line. $\R=5772$, $m=0.7$, $p=0.6$, $q=0.4$,
$\Gamma_{p+}= -0.0345$, $\Gamma_{p-}=0.0138$. }
 \label{f:polyprim}
\end{figure}

One could think that the strong stabilization discussed in the
previous subsection is only due to the precise positioning of the
mixed layer at the critical layer. If so, the result would have very
little generic consequence.  In this subsection we show that any
reasonable viscosity profile achieves the same effects. To this aim we
consider the effective viscosity profile reported in \cite{97SBH} (in
their Fig. 5) which is obtained from simulations of a turbulent
channel flow with polymer additive. It may be prescribed as
\begin{eqnarray}\label{cmuin}
\mu(y) &=& 1
\,,\quad  \mbox{for} \quad 0 \le y \le p\,,\\
\label{vis3}
\mu(y) &=& 1 +
(m-1)\, \left({y-p \over q}\right)^3,
\end{eqnarray}
with $q \sim 0.4$, and
$m \sim 0.7$, as shown in Fig.
\ref{f:contprof}.
The energy balance for the least stable primary mode at $\R=5772$ for
this case (Fig. \ref{f:polyprim}) shows a large counter-production of
disturbance kinetic energy, which is in fact more pronounced than what
we obtained with a mixed layer (Fig. \ref{f:near}). Thus the strong
stabilization effect does not require careful placing of the viscosity
variation at a particular layer.  It is sufficient that there exist a
viscosity variation in the region of the critical layer (indicated as
the vertical line in Fig. \ref{f:polyprim}) to achieve the
stabilization.

It comes as no surprise that this continuous viscosity profile behaves
very similarly to the thin mixed-layer. If we return to equation
(\ref{critlow}), we will see that all we have now done is to increase
both $\nu'$ (which is proportional to $m-1$) and $q$ threefold (the
effective $q$ here is closer to 0.3 than 0.4, as we can see from
Fig. \ref{f:contprof}), so the ratio remains the same.

\section{Secondary Instabilities}
\label{secondary}  

A laminar flow through a channel is linearly unstable at $\R=5772$. In
all except the cleanest experiments, however, the flow becomes
turbulent at much lower Reynolds numbers, as low as $1000$
\cite{80OK,93TTRD}. This is because the linear stability analysis is
carried out on a steady laminar velocity profile, whereas a real flow,
except under carefully designed clean conditions, consists in addition
of small but finite disturbances (most of whom will decay at long
times). The stability behaviour of the real flow is quite different
from that of the steady profile: the actual flow is unstable to new
modes, often referred to as secondary modes. The secondary modes are
often three dimensional, and their signature is prominent in
fully-developed turbulence. As described below, the secondary
instabilities are studied by a Floquet analysis of the periodic
primary flow we obtained earlier.

As is usual in the analysis of secondary instabilities
\cite{88Herb,83Herb}, we begin by splitting the flow into a periodic
component (consisting of the mean laminar profile in addition to the
primary wave) and a secondary disturbance, e.g., \be \B.U_{\rm
total}(x,y,z,t) = \B.U(x,y,t) + \B.u_s(x,y,z,t), \ee where
\begin{eqnarray}
&&\!\!\B.U(x,y,t) = \bar U(y) \hat \B.x \label{basic}\\&&\!\!+ A_p(t)
\left\{\left[u_p(y) \hat \B.x  + v_p(y) \hat \B.y
\right] \exp \left[\i k_p (x - c_p t)\right] + {\rm c.c.} \right\}.
\nonumber
\end{eqnarray}
Here $\hat \B.x$ and $\hat \B.y$ are units vectors in the $x$
(steamwise) and $y$ (wall normal) directions.  The amplitude $A_p$ of
the primary disturbance changes very slowly with time, and $\d A_p /\d
t$ may be neglected during one time period. The spatial and temporal
dependence of the secondary disturbance is written in the form
\begin{eqnarray}
&&\B.u_s(y,\B.r_\perp,t)\equiv {\cal R}\!e\Big\{\B.u_{s+}(y) \exp\left[\i
\left(\B.k_+ \cdot \B.r_\perp - \omega_+ t\right)\right] \nonumber\\&&+
\B.u_{s-}(y) \exp\left[\i \left(\B.k_-\cdot \B.r_\perp - \omega_- t\right)
\right]\Big\},
\label{form}
\end{eqnarray}
where $\B.r_\perp\equiv x\hat \B.x +z\hat \B.z $, and $\B.k_\pm= k_\pm
\hat\B.x \pm k_z \hat\B.z$.  We substitute the above ansatz into the
Navier-Stokes and continuity equations, and retain linear terms in the
secondary. On averaging over $x$, $z$ and $t$, only the resonant modes
survive, which are related by \be \B.k_+ + \B.k_- = k_p \hat \B.x \ ,
\quad \text{therefore}~~\B.k_\pm=\pm \B.q+\frac{k_p}{2} \hat \B.x \ ,
\ee for any vector $\B.q$, and \be \omega_+ = \omega + \i\gamma_s
\qquad {\rm and} \qquad \omega_- = (\omega_p - \omega) + \i\gamma_s.
\label{om}
\ee Eliminating the disturbance pressure and streamwise component of
the velocity, we get the equations for the secondary disturbances
$v_s$ and $w_s$.  Using the operator $D$ for differentiation with
respect to the normal coordinate $y$, and the notation $f_\pm \equiv -
\i w_{s\pm}/k_z$, the equations read \cite{02SG}
\begin{eqnarray}
&&\A \nonumber\\&&
\times\left[\s f_+ - Dv_+\right] - \i k_+ U' v_+ \label{first}\\&&-
{A_p k_+ \over 2 k_-}\Bigg\{\left[\i k_+ u_p D + v_p D^2 +\i k_- Du_p
\right]v_-^*
\nonumber\\&&+ \left[\left(\beta^2-k_-k_+\right)v_pD + \i k_+
\left(k_-^2+\beta^2\right)u_p\right]f_-^*\Bigg\} = 0,
\nonumber
\end{eqnarray}
and
\begin{eqnarray}
&&\A 
\left(Df_+ - v_+\right) \nonumber\\&&+ \bigg[-\i k_+ (DU) + (D^2\nu) D +
(D\nu)(D^2 - k_+^2 - \beta^2)\bigg] f_+ \nonumber\\&&+ {A_p (k_p + k_-)
\over 2} \bigg[ \i u_p \left(v_-^* + Df_-^*\right)
-{v_p \over k_-}Dv_-^* \bigg] \nonumber\\&&+
{A_p \over 2}\left[v_p\left({k_p\beta^2 \over k_-} + D^2\right) -
\i k_-(Du_p) \right]f_-^* = 0,
\label{second}
\end{eqnarray}
The boundary conditions are
\be
\B.u_s=0 \qquad {\rm at} \qquad y = \pm 1.
\label{bcs}
\ee Equations (\ref{first}) and (\ref{second}), along with two
corresponding equations in $v_-^*$ and $f_-^*$, describe an eigenvalue
problem for the secondary instability. The four equations are solved
by a Chebychev collocation spectral method, details of the solution
procedure are available in \cite{02SG}.

The most unstable secondary mode in our case is found to be the
subharmonic, for which $\B.q=\beta \hat \B.z$.  The production and
dissipation are computed as before.

We survey in turn the thin mixed-layer profile, and the continuous
viscosity profile to see what viscosity variation does to the
secondary instability.
\subsection{Mixed Layer}
\label{thin-sec}

\begin{figure}
\epsfxsize=7.5cm \epsfbox{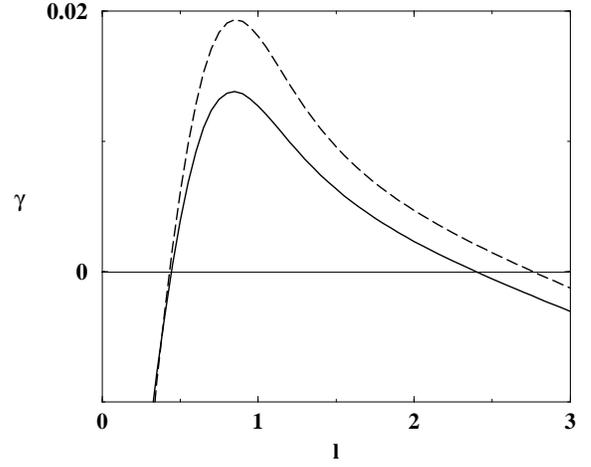}
\caption{Dependence of growth rate on spanwise wavenumber. Solid line:
varying viscosity ($p=0.8, q=0.1, m=0.9$); dashed line: constant
viscosity ($m=1$). $k_p=1, A_p=0.005, \R=6000$. }
\label{f:thinmax}
\end{figure}

\begin{figure}
\epsfxsize=7.5cm \epsfbox{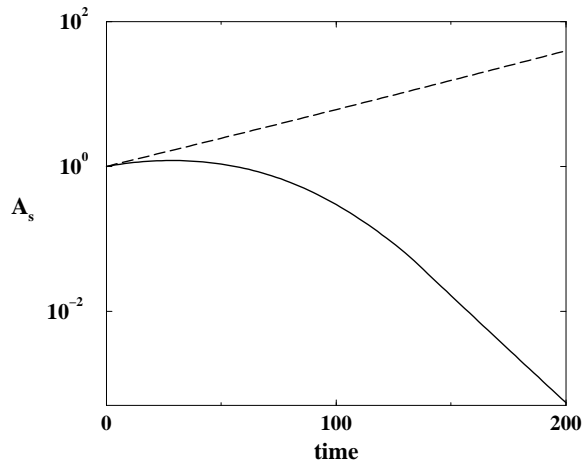}
\caption{Amplitude of the secondary mode in logarithmic 
scale as a function of time. Dashed line: constant viscosity,
$m=1$. Here $\gamma_p=0.0003$, and the primary mode is unstable. Solid
line: varying viscosity; here $\gamma_p=-0.0206$, the primary mode is
stable. All conditions like in Fig. \ref{f:thinmax}, in particular
$A_p(t=0)=0.005$.}
\label{f:thinamp}
\end{figure}

The velocity and viscosity profiles here are as given in Fig.
\ref{f:thinprof}, and the primary instability is that presented in
Sect.  \ref{primix}. Since the subharmonic ($k_+=k_-= k_p/2$) is the
least stable mode, we present this case alone. In Fig. \ref{f:thinmax}
a typical dependence of the growth rate of the secondary mode on the
spanwise wavenumber is shown. We can see that the viscosity variation
damps the secondary mode significantly, but it is still
unstable. However, there is a crucial difference in the {\em primary}
instabilities of the two: the primary is unstable for a constant
viscosity flow, but very stable in the mixed layer case. Therefore at
long times, the secondary mode, which feeds on the primary for its
existence, dies down in the latter case.  To compute the time
dependence of the amplitude of the secondary mode we computed the
growth rate $\gamma_s$ by neglecting the time dependence of the
amplitude of the primary words.  As a result we obtain the growth rate
$\gamma_s[A_p(t)]$, in which $A_p(t)$ can be an exponentially growing
or a decaying function of time. Having this growth rate we can present
the time dependence of the amplitude of the secondary mode, see Fig.
\ref{f:thinamp}. Without the viscosity contrast, the amplitude of the
secondary mode increases (essentially exponentially).  With the
viscosity contrast the amplitude decays in time.

\begin{figure}
\epsfxsize=7.5cm \epsfbox{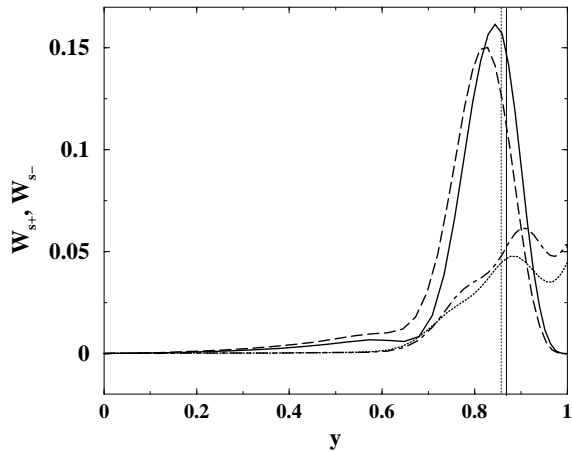}
\caption{Production $W_{s+}$ and dissipation $W_{s-}$ of the 
kinetic energy
of the secondary disturbance at time=0. Solid line: $W_{s+}, m=0.9$;
dot-dashed line: $W_{s-}, m=0.9$; long dashes: $W_{s+}, m=1$; dotted
line: $W_{s-}, m=1$.  The vertical lines show $y_c$ (the critical
point location) for $m=0.9$ (solid) and $m=1$ (dotted).}
\label{f:thinprod0}
\end{figure}

\begin{figure}
\epsfxsize=7.5cm \epsfbox{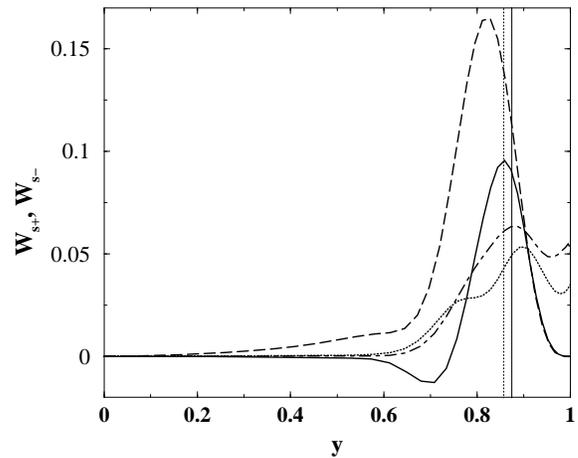}
\caption{Production $W_{s+}$ and dissipation $W_{s-}$ of 
the kinetic energy of the secondary disturbance at time=40. Solid
line: $W_{s+}, m=0.9$; dot-dashed line: $W_{s-}, m=0.9, A_p=0.00215$;
long dashes: $W_{s+}, m=1$; dotted line: $W_{s-}, m=1, A_p=0.00506$.}
\label{f:thinprod40}
\end{figure}

We now observe the balances of energy initially and at a later time in
Figs.  \ref{f:thinprod0} and \ref{f:thinprod40} respectively. The
initial balance of energy is not very different from the constant
viscosity case.  At the later time, however, the production of
secondary kinetic energy is significantly lower. The location $y_c$ of
the critical point is seen from the figures to be close to the layer
of stratified viscosity. If the two were well-separated, the
stratification would do nothing to the secondary mode.

A lowest-order analysis of the secondary stability equations is not as
straightforward as for the primary mode, since the secondary is highly
dependent on the amplitude of the primary \cite{02SG}. We may however
make the following observations from a critical layer analysis of
equations (\ref{first}) and (\ref{second}) and their
counterparts. When $A_p\gg \ep$, (cf. Eq.(\ref{epsilon})) only the
nonlinear terms appear at the lowest order, and the secondary mode is
completely driven by the primary.  When $A_p \sim O(\ep)$, both the
basic terms and the nonlinear terms contribute at the lowest order. It
may be numerically determined, however, that the secondary is slaved
to the primary here as well. When $A_p=o(\ep)$, the lowest-order
theory for the secondary is (not surprisingly) exactly that given by
(\ref{critlow}) for the primary.

A direct estimate of the effect of the viscosity stratification on the
secondary mode is obtained from the threshold amplitude $A_{\rm th}$ of
the primary for the instability. At a Reynolds number of 6000 and
primary wavenumber of $k_p=1$, for a neat fluid, all secondary modes
are damped if $A_{\rm th}<0.002$, while for the continuous viscosity
profile, all secondary modes continue to be damped even for larger
primary disturbances, up to $A_{\rm th}=0.005$. When the Reynolds number
is reduced to 2000, the threshold amplitudes are 0.012 and 0.016 for
the neat and viscosity-stratified fluids respectively.
\subsection{Continuous Viscosity Profile}
\label{cont-sec}

\begin{figure}
\epsfxsize=7.5cm \epsfbox{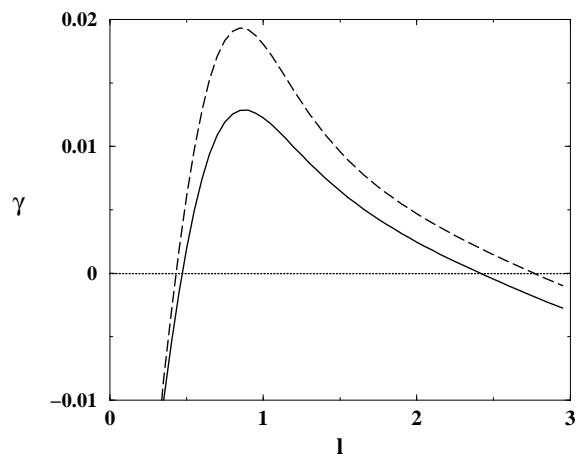}
\caption{Dependence of growth rate on spanwise wavenumber. Solid line:
varying viscosity [according to equation (\ref{vis3})]; dashed line:
constant viscosity. Wavenumbers and $\R$ as in Fig. \ref{f:thinmax}.}
\label{f:max}
\end{figure}

\begin{figure}
\epsfxsize=7.5cm \epsfbox{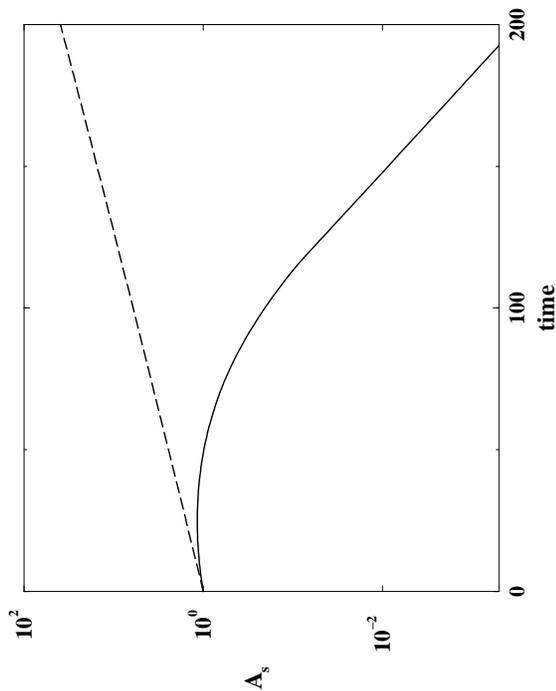}
\caption{Variation of the amplitude of the secondary instability mode with
time. Solid line: varying viscosity, $\gamma_p=-0.0244$; dashed line: 
constant viscosity, $\gamma_p=0.0003$.
Wavenumbers and $\R$ as in Fig. \ref{f:thinamp}.}
\label{f:amp}
\end{figure}

\begin{figure}
\epsfxsize=7.5cm \epsfbox{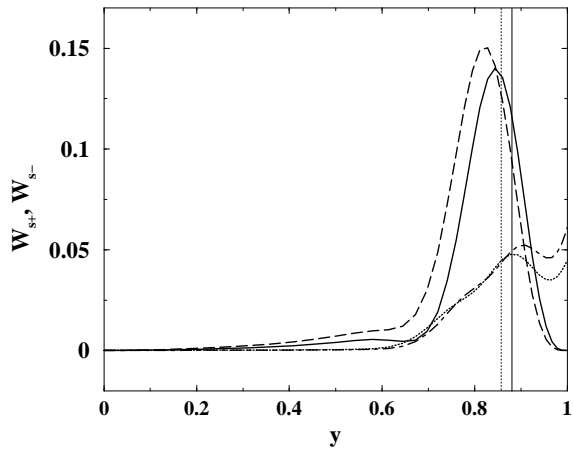}
\caption{Production and dissipation at time=0. Solid line: varying
viscosity;
dashed line: constant viscosity, $A_p=0.005$ for both.}
\label{f:prod0}
\end{figure}

\begin{figure}
\epsfxsize=7.5cm \epsfbox{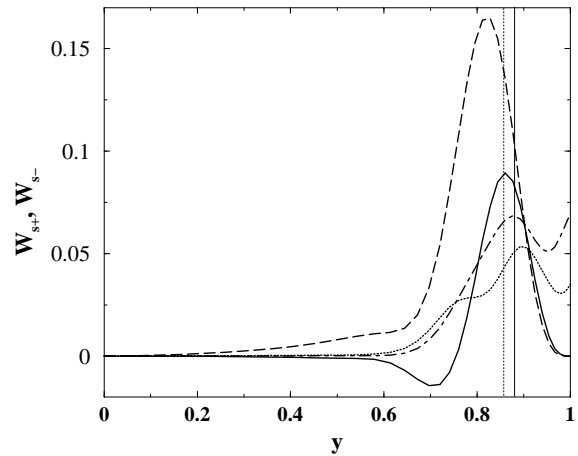}
\caption{Production and dissipation at time=40. Solid line: varying
viscosity, $A_p=0.00184$;
dashed line: constant viscosity, $A_p=0.00506$.}
\label{f:prod40}
\end{figure}

The velocity and viscosity profiles here are as given in Fig.
\ref{f:contprof}, and the primary instability is that presented in Sect.
\ref{contprim}. The counterparts for the continuous viscosity
profile of Figs. \ref{f:thinmax} to \ref{f:thinprod40} are presented
in Figs. \ref{f:max} to \ref{f:prod40} respectively. It is clear that
nothing has changed qualitatively.

\begin{figure}
\epsfxsize=7.5cm \epsfbox{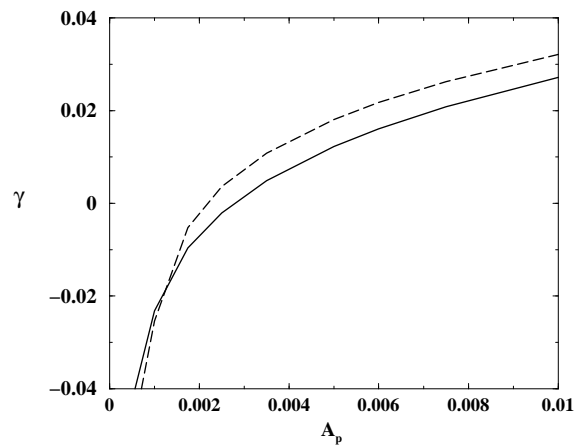}
\caption{Dependence of the growth rate of the secondary mode on the
amplitude of the primary disturbance. Solid line: stratified 
viscosity; dashed line: constant viscosity. $k_p=1, k_+=0.5, \beta=1,
\R=6000$.}
\label{f:ampdep}
\end{figure}

Fig. \ref{f:ampdep} shows the dependence of the growth rate of the 
secondary mode on the amplitude of the primary disturbance. It is
clear that the instability is reduced by the stratification of
viscosity, but there is no dramatic effect in the secondary alone. We
may conclude that the large effect comes from the complete reliance 
of the secondary on the primary.
\section{Concluding Remarks}
\label{conclude}

We addressed the primary and secondary instability of simple channel
flows, and examined the effects of small viscosity variations. We find
dramatic effects of stabilization when the viscosity variations exist
in the vicinity of the critical layers, in which the speed of
propagation of the modes coincided with the mean velocity of the basic
flow. With about 10\% viscosity changes we can have very large
increases in the threshold Reynolds numbers for instability. In all
cases we find that the main mechanism for the large effects is the
reduction of the intake of energy from the mean flow to the putative
unstabale modes, which therefore become stable. For the same Reynolds
numbers in Newtonian fluids there is no such mechanism for
stabilization and these flows will become turbulent. We would like to
propose that similar effects should be examined in the case of
turbulent drag reduction by polymer additives.

We recognize that in a turbulent flow there are many more modes that
interact, but we propose that a similar mechanism operates for each
mode at its critical layer, where both elastic and viscous effects
determine the mean flow. The advantage of the present calculation is
that we can consider all the putative unstable modes, and conclude
that with a viscosity gradient similar to that seen in polymeric
turbulent flows the linear threshold $\R_{\rm th}$ goes up five times
(to 31000). We note in passing that this effect had not been put to an
experimental test, and it would be exciting to have a confirmation of
our predictions by future experiments. For actual turbulent flows we
will need first to identify what are the main modes that interact
between themselves and with the mean flow. A significant numerical
effort is required, but appears worthwhile due to the importance of
the phenomenon of drag reduction, and its relative lack of
understanding.

We have demonstrated that the exact form of the viscosity profile is
immaterial; a continuous profile of viscosity in the critical region
behaves exaclty like a thin mixed layer. We have shown that the secondary
three dimensional modes of instability are ``slaved'' to the primary
linear mode of instability: the mechanism which stabilizes the primary
mode indirectly ensures that the secondary is damped out quickly.

Finally we note that a linear disturbance can rear its head either in
the form of the fastest growing (or slowest decaying) mode as
considered here; or in non-modal form with a transient growth followed
by long-term decay \cite{02Chap}. The former situation will correspond
to relatively high Reynolds numbers, or cleaner set-ups. We expect
similar conclusions in the latter situation as well.
\acknowledgments This work was supported by the German-Israeli
Foundation, the European Commission under a TMR grant, the Israeli
Science Foundation and the Naftali and Anna Backenroth-Bronicki Fund
for Research in Chaos and Complexity.  RG thanks the Defence R\&D
Organisation, India, for financial support.


\end{document}